\documentclass[reprint,showpacs,preprintnumbers,amsmath,amssymb,prl]{revtex4}

\usepackage{amsmath}
\usepackage{amsfonts}
\usepackage{amsthm}
\usepackage{dsfont}
\usepackage{graphics}
\usepackage{graphicx}
\usepackage{subfigure}
\usepackage{amssymb}
\usepackage {mathrsfs}
\usepackage{braket}
\usepackage{bbold}
\usepackage[bb=boondox]{mathalfa}
\usepackage{bbm}
\usepackage{color}

\begin{document}
\title{Relativistic Wigner function for quantum walks}
\author{Fabrice Debbasch}
\email{fabrice.debbasch@gmail.com}
\affiliation{Sorbonne Universit\'e, Observatoire de Paris, Universit\'e PSL, CNRS, LERMA, F-75005, Paris, France}
\date{\today}

%
%





\begin{abstract}

A relativistic Wigner function for free Discrete Time Quantum Walks (DTQWs) on the square $2D$ space-time lattice is defined. Useful concepts such as discrete derivatives and discrete distributions 
are also introduced. The transport equation obeyed by the relativistic Wigner function is obtained and degenerates at the continuous limit into the 
transport equation obeyed by the Wigner function of $2D$ Dirac fermions. The first corrections to the continuous equation induced by the discreteness of the lattice are also computed.

\end{abstract}


\maketitle

\section{Introduction}

Discrete Time Quantum Walks (DTQWs) are unitary quantum automata that can be viewed as formal generalisations of classical random walks. Following the seminal work of Feynman \cite{FeynHibbs65a}
Gr\"ossing and Zeilinger \cite{GZ88a} \color{black} and Aharonov \cite{ADZ93a} they were considered in a systematic way by Meyer \cite{Meyer96a}. DTQWs have been realized experimentally with a wide range of physical objects and setups \cite{Schmitz09a, Zahring10a, Schreiber10a, MA13a,Karski09a, Sansoni11a, Sanders03a, Perets08a}, and are studied in a large variety of contexts, ranging from quantum optics \cite{Perets08a}
to quantum algorithmics \cite{Amb07a, MNRS07a}, condensed matter physics \cite{Aslangul05a, Bose03a, Burg06a, Bose07a,DFB15a}, hydrodynamics \cite{HMDB19a} and biophysics \cite{Collini10a, Engel07a}.

It is well known that several DTQWs can be viewed as discrete versions of the continuous Dirac 
and Weyl \color{black} fermion dynamics 
\cite{BB94a,Y05a,Strauch06a,Strauch07a,ANF14a,BDAT15a,BDAPT15a,DAEP17a}. 
\color{black}
These DTQWs have a continuous limit which coincides with the Dirac equation and they even display exact discrete gauge invariance properties \cite{CW13a,DBD13a,DBD14a,
AF16b,AMBD16a}.
What however remains unknown is the phase-space behaviour of these DTQWs. If one follows the standard procedure adopted for Dirac 
fermions \cite{Elze86a,Vasak87a,Gao17a}, one should first build a relativistic Wigner function for DTQWs and then describe the phase-space dynamics by the transport equation obeyed by that function. Until now, the only Wigner functions \color{black}  that have \color{black} been considered for DTQWs 
\cite{Perez12a,HBP13a,HBP15a,AWM14a} are \color{black} non relativistic \cite{Curt14a}. Thus, 
these \color{black} function
and the equation 
they \color{black} obey do not coincide, at the continuous limit, with the usual Wigner function and phase-space transport equation for Dirac fermions. 

The aim of this article is to fill this gap for free DTQWs in flat $2D$ space-time. We first generalize basic concepts of continuous mathematics such as derivation and distribution theory to analysis on discrete lattice. We then define a discrete relativistic Wigner function for DTQWs and derive the corresponding discrete relativistic transport equation. At the continuous limit, the discrete Wigner function tends towards the Wigner function of Dirac particles and the transport equation tends towards the equation obeyed by the Wigner function of Dirac fermions
in continuous space-time. We also compute the first correction to the transport equation induced by the discreteness of the lattice on which the DTQWs propagate. We finally discuss all results in the last section of the article.




\section{\label{sec:basics}A simple Dirac QW }

We work with two-component wave-functions $\Psi$
defined in $2D$ discrete space-time where instants are labeled by 
$j \in \mathbb{Z}$,
spatial positions are labeled by 
$p \in \mathbb{Z}$
 and 
 $\Psi_j = (\psi_{j, p})_{p \in \mathbb Z}$. 
 We introduce a basis $(b_A) = (b_L, b_R)$ in Hilbert-space space and the components $\Psi^A = (\Psi^L, \Psi^R)$ of the arbitrary wave-function $\Psi$ in this basis. The Hilbert product is defined by $<\psi, \phi> = \sum_{A, j, p} (\psi^A)^*_{j, p}  (\phi^A)_{j, p}$, which makes the basis $(b_A)$ orthonormal. 
Consider now the quantum walk $\Psi_{j + 1} = U_j T \Psi_j$
where $T$ is the spatial-translation operator defined by $(T \Psi_j)_{j, p} 
= (\psi^L_{j, p+1}, \psi^R_{j, p-1})^T$ and $U_j$ is an $SU(2)$ operator defined by 
\begin{equation}
(U_j \Psi_j)_{j, p} = U (\theta) \psi_{j, p}
\end{equation}
where
\begin{equation}
U(\theta) = 
\left( \begin{array}{cc}  \cos \theta & -i \sin \theta \\
- i \sin \theta&  \cos \theta \end{array} \right),
\end{equation}
with constant $\theta$.
It was shown in \cite{DBD14a} that this quantum walk, at the continuous limit defined by $t_j = j \epsilon$, $x_p = p \epsilon$, $\theta = \epsilon m$, tends to the Dirac equation
for a $2$D spinor of mass $m$ in flat Minkovski space-time with coordinates $(t, x)$ as $\epsilon$ tends to zero.

%
%

\section{Basic tools}
\subsection{Discrete derivatives}

Let us now rewrite the above equations with the help of discrete covariant derivative.
We define:
%
%
%
\begin{eqnarray}
(D_j f)_{j, p} & = & \frac{1}{2}\, (f_{j+1,p} - f_{j-1,p}) \nonumber \\
(D_{jj} f)_{j, p} & = & \frac{1}{2} \left( 
f_{j+1, p} + f_{j-1, p} - 2 f_{j,p}\right) \nonumber \\
(D_p f)_{j, p} & = & \frac{1}{2} \left( 
f_{j, p+1} - f_{j, p-1}\right), \nonumber \\
(D_{pp} f)_{j, p} & = & \frac{1}{2} \left( 
f_{j, p+1} + f_{j, p-1} - 2 f_{j,p}\right)
\end{eqnarray}
where $f$ is an arbitrary $j$- and $p$-dependent quantity. These are 
discrete versions of the usual partial derivatives. Inverting the above equations delivers:
%
%
%
%
\begin{eqnarray}
f_{j+1,p} & = & f_{j,p} +  (D_j f)_{j, p} +  (D_{jj} f)_{j, p}\nonumber \\
f_{j-1,p} & = & f_{j,p} -  (D_j f)_{j, p} +  (D_{jj} f)_{j, p}\nonumber \\
f_{j, p+1} & = & f_{j,p} +  (D_p f)_{j, p} +  (D_{pp} f)_{j, p} \nonumber \\
f_{j, p-1} & = & f_{j,p} -  (D_p f)_{j, p} +  (D_{pp} f)_{j, p}.
\label{eq:deruseful}
\end{eqnarray}

The equation of motion of the QW can then be rewritten as:
\begin{equation}
(D_j \Psi^A)_{j,p}  =  (U \sigma_3)^A_B (D_p \Psi^B)_{j, p} 
+  
(U - \mathbbm{1})^A_B \Psi^B_{j, p} 
\color{black}
 -  \delta^A_B (D_{jj} \Psi^B)_{j, p} + U^A_B (D_{pp} \Psi^B)_{j, p} \ , 
\label{eq:FinDiffII}
\end{equation}
where
$\sigma_3$ is the operator represented by the third Pauli matrix in the basis $(b_A)$ {\sl i.e.} 
$\sigma_3$ is represented by the matrix $\mbox{diag} (1, -1)$ in the basis $(b_A)$
and $(\mathbbm{1})^A_B = \delta^A_B = 1$ if $A = B$ and $0$ otherwise.
At the continuous limit, the left-hand side and the first term on the right-hand side deliver 
the differential terms in the Dirac equation, the second contribution to the right-hand side delivers the mass term
while the other two terms vanish because they are of higher order.

Let us finally mention the following identities, which will be used in the next sections:
\begin{equation}
D_j (f g)  =  (D_j f) g + f (D_j g) 
 +  (D_j f) (D_{jj} g)  + (D_{jj} f) (D_j g),
\label{eq:Djproduct}
\end{equation} 
\begin{equation}
D_p (f g) =  (D_p f) g + f (D_p g) 
 + (D_p f) (D_{pp} g)  + (D_{pp} f) (D_p g).
\label{eq:Dpproduct}
\end{equation}
%
Note that the terms which vanish at the continuous limit are of order $3$, and not $2$ in $\epsilon$. 

\subsection{Discrete distributions}

In what follows, we will consider discrete Fourier transforms of quantities which do vanish at infinity. To give meaning to these Fourier transforms, one has to extend the 
theory of distributions to the discrete case. 
For simplicity sakes, all ideas are now presented for functions and distributions of one single discrete variable $n\in {\mathbb Z}$, which plays to role of $j$ or $p$. The extension to two discrete variables used in the following sections is straightforward.
\color{black}

We introduce as test 
functions the space of all functions $h$ of $n$ which admit a discrete Fourier transform ${\hat h}$ defined by 
\begin{equation}
{\hat h}(k) = 
\sum_n \exp (i k n) h_n.
\end{equation}
All these functions vanish at infinity. The conjugate momentum $k$ takes value in the first Brillouin zone of the lattice {\sl i.e.} $(- \pi, \pi)$.
The inverse Fourier transform is thus defined by:
\begin{equation}
{h}_n = 
\frac{1}{2 \pi}\, 
\int_{- \pi}^{\pi} 
\exp ( - i k n) 
{\hat h} (k) dk.
\end{equation}

Any function $f$ defined on $\mathbb Z$ can now be considered a distribution acting on this space of test functions and we define
\begin{equation}
f(h) = 
<f, h >_n 
\color{black}
= \sum_n f_n h_n.
\end{equation}
We now define the discrete Fourier transform $\hat f$ of the distribution $f$ by its
action on functions on the variable $k$:
\begin{eqnarray}
{\hat f} (H) & = & 
\sum_{n}
\frac{1}{2 \pi} \int_{- \pi}^{\pi} dk \exp (i k n) f_n H(k) \nonumber \\
& = &  <f , {\hat H} >_n
\end{eqnarray}
\color{black}
Introducing the natural product in Fourier-space
\begin{eqnarray}
< u, v>_k = \frac{1}{2 \pi} \int_{- \pi}^{\pi} dk \,  u(k) v(k),
\end{eqnarray}
the definition of ${\hat f}$ can be rewritten as $< {\hat f}, H >_k = < f, {\hat H}>_n$.
\color{black}

Consider the Fourier transform of the derivative $F = D_nf$. 
\begin{eqnarray}
<{\hat F}, H >_k & = &
 <F , {\hat H} >_n  \nonumber \\
& = & \sum_{n}
\frac{1}{2 \pi} \int_{- \pi}^{\pi} dk \exp (i k n) F_n H(k) \nonumber \\
& = & \sum_{n}
\frac{1}{2 \pi} \int_{- \pi}^{\pi} dk \exp (i k n) \frac{1}{2} \left(
f_{n+1} - f_{n-1}
\right) H(k) \nonumber \\
& = & \sum_{n}
\frac{1}{2 \pi} \int_{- \pi}^{\pi} dk \exp (i k n) \frac{1}{2} \left(
\exp(- i k) - \exp(ik)
\right) f_n H(k) \nonumber \\
& = & \sum_{n}
\frac{1}{2 \pi} \int_{- \pi}^{\pi} dk \exp (i k n) f_n \left(
- i \sin k 
\right)  H(k) \nonumber \\
& = & < G, H >_k
\end{eqnarray}
where $G(k) = - i (\sin k) {\hat f}(k)$. 
Thus, 
\begin{eqnarray}
{\hat F} (k) = - i (\sin k)  {\hat f} (k). \label{eq:FourierDer}
\end{eqnarray}
\color{black}
%
%
At the continuous limit, the continuous position $N$ and wave-vector $K$ are related to $n$ and $k$ by $N = \epsilon n$ and $K = k/\epsilon$, so (\ref{eq:FourierDer})
becomes
\begin{eqnarray}
\epsilon {\widehat 
{f'}
} (K)  = - i \sin (\epsilon K)  {\hat f} (K) 
\label{eq:FourierDer1}
\end{eqnarray}
\color{black}
where $f' = df/dN$. 
At orders $0$ and $1$, (\ref{eq:FourierDer1}) reads simply
\begin{eqnarray}
{\widehat 
{f'}
} (K)  = - i K {\hat f} (K) \label{eq:FourierDer1bis}
\end{eqnarray}
but one finds, for example at second order in $\epsilon$:
\begin{eqnarray}
{\widehat 
{f'}
} (K)  = - i K \left( 1 -  
\frac{\epsilon^2 K^2}{3}
\color{black}
 \right) {\hat f} (K).
\label{eq:FourierDer1ter}
\end{eqnarray}
\color{black}






\section{Discrete Wigner function}

We now introduce the `density' in space-time $\Omega^{AB}_{j, p, n_j, n_p} = (\Psi^A)^*_{j - n_j, p - n_p} (\Psi^B)_{j + n_j, p + n_p}\ $, $(j, n_j, p, n_p) \in {\mathbb Z}^4$, $(A, B) \in
\left\{L, R\right\}^2$. We consider that this object is, at fixed $(j, p)$, a discrete distribution acting on functions of $(n_j, n_p)$ which admit a discrete Fourier transform with respect to these variables and we define the discrete Wigner `function' $W^{AB}$ as the Fourier transform of $\Omega^{AB}$ with respect to $(n_j, n_p)$:
\begin{equation}
W^{AB}_{j, p, k_j, k_p}  
\color{black}
=  \sum_{(n_j, n_p) \in {{\mathbb Z}^2}} \exp{\left(+ i k_j n_j + i k_p n_p \right)} 
 \times 
(\Psi^A)^*_{j - n_j, p - n_p} (\Psi^B)_{j + n_j, p + n_p}.
\color{black}
\end{equation}

%
%
%
Let us now obtain from the equations of motion of the QW an equation of motion for $W$.
Following the computation carried out in the continuous case {\sl i.e.} for the Dirac equation, we first compute
the discrete derivatives of $\Omega^{AB}$
with respect to $j$ ($D_j$), $p$ ($D_p$), $n_j$ ($D_{n_j}$) and $n_p$ ($D_{n_p}$), where the derivatives $D_{n_j}$ and $D_{n_p}$ are defined as
$D_j$ and $D_p$ above. The derivatives of $\Omega$ are best computed using the identities (\ref{eq:Djproduct},\ref{eq:Dpproduct}).
One obtains
\begin{equation}
(D_j \Omega^{AB})_{j, p, n_j, n_p}   =  (D_j  (\Psi^A)^*)_{j - n_j, p - n_p} (\Psi^B)_{j + n_j, p + n_p}
+  (\Psi^A)^* _{j - n_j, p - n_p} (D_j \Psi^B)_{j + n_j, p + n_p} 
 +  (\Delta_j^{AB})_{j, p, n_j, n_p} \ ,
\end{equation}
\begin{equation}
(D_{n_j} \Omega^{AB})_{j, p, n_j, n_p}  
\color{black} =  
- 
\color{black}
(D_j  (\Psi^A)^*)_{j - n_j, p - n_p} (\Psi^B)_{j + n_j, p + n_p}
+  (\Psi^A)^* _{j - n_j, p - n_p} (D_j \Psi^B)_{j + n_j, p + n_p} 
 +  
 (\Delta_{n_j}^{AB})_{j, p, n_j, n_p} \ ,
\end{equation}
\color{black}
where
\begin{equation}
(\Delta_j^{AB})_{j, p, n_j, n_p}  =  (D_j  (\Psi^A)^*)_{j - n_j, p - n_p} (D_{jj} \Psi^B)_{j + n_j, p + n_p} 
+  (D_{jj} (\Psi^A)^*)_{j - n_j, p - n_p} 
(D_{j} \Psi^B)_{j + n_j, p + n_p} \ ,
\end{equation}
\begin{equation}
(\Delta_{n_j}^{AB})_{j, p, k_\alpha, k_\beta}  =  - (D_j  (\Psi^A)^*)_{j - n_j, p - n_p} (D_{jj} \Psi^B)_{j + n_j, p + n_p} 
+ 
\color{black}
(D_{jj} (\Psi^A)^*)_{j - n_j, p - n_p} (D_{j} \Psi^B)_{j + n_j, p + n_p}.
\end{equation}
\color{black}
Similar relations can be written for $D_p$ and $D_{n_p}$. 

Putting all this together delivers
\begin{eqnarray}
\left( (D_j + D_{n_j}) \Omega^{AB}\right)_{j, p, n_j, n_p}  & = & 2 (\Psi^A)^*_{j - n_j, p - n_p} (D_j \Psi^B)_{j + n_j, p + n_p} 
 + 2 (D_{jj} \Psi^A)^*_{j - n_j, p - n_p} (D_j \Psi^B)_{j + n_j, p + n_p} 
 \color{black}
\ ,
\end{eqnarray}
and
\begin{eqnarray}
\left( (D_p + D_{n_p}) \Omega^{AB}\right)_{j, p, n_j, n_p}  & = & 2 (\Psi^A)^*_{j - n_j, p - n_p} (D_p \Psi^B)_{j + n_j, p + n_p}
 + 2 (D_{pp} \Psi^A)^*_{j - n_j, p - n_p} (D_p \Psi^B)_{j + n_j, p + n_p} 
 \color{black}
.
\end{eqnarray}

Using the equation of motion (\ref{eq:FinDiffII}) leads to 
\begin{eqnarray}
& & \left( (D_j + D_{n_j}) \Omega^{AB}\right)_{j, p, n_j, n_p}  - (U \sigma_3)^B_C \left( (D_p + D_{n_p \color{black}}) \Omega^{AC}\right)_{j, p, n_j, n_p}  = \nonumber \\
& & 
2 
\color{black}
(\Psi^A)^*_{j - n_j, p - n_p} (U - \mathbbm{1})^B_C (\Psi^C)_{j + n_j, p + n_p}
 - 2 (\Psi^A)^*_{j - n_j, p - n_p} \left( \left( \delta^B_C D_{jj}  -  U^B_C D_{pp} \right) (\Psi^C) \right)_{j + n_j, p + n_p} \nonumber \\
 & + & 
 2 \left( D_{jj}(\Psi^A)^*\right)_{j - n_j, p - n_p} \left(D_j (\Psi^B) \right)_{j + n_j, p + n_p} 
 - 2 \left( D_{pp}(\Psi^A)^*\right)_{j - n_j, p - n_p} (U \sigma_3)^B_C \left(D_j (\Psi^C) \right)_{j + n_j, p + n_p}.
\color{black}
\end{eqnarray}

Taking the Fourier transform delivers:
\begin{eqnarray}
\left(  \delta^B_C D_j   - (U \sigma_3)^B_C D_p \right) W^{AC} & = & K^{AB}\left[ \Omega \right] + M^{AB} \left[ \Psi^*, \Psi \right] 
\end{eqnarray}
where
\begin{equation}
\left( K^{AB}\left[ \Omega \right] \right)_{j, p, k_j, k_p}  =  \sum_{(n_j, n_p) \in {{\mathbb Z}^2}} \exp{\left(+ i k_j n_j + i  k_p n_p \right)}
 \times  
\left( \delta^B_C D_{n_j}  - (U \sigma_3)^B_C D_{n_p} \right) \Omega^{AC}_{j, p, n_j, n_p}
\end{equation}
and
\begin{equation}
 \left( M^{AB}\left[  \Psi^*, \Psi \right] \right)_{j, p, k_j, k_p}  = \left( M^{AB}_c\left( W\right) \right)_{j, p, k_j, k_p} + \left( M^{AB}_s\left[  \Psi^*, \Psi \right] \right)_{j, p, k_j, k_p}
\end{equation}
with
\color{black}
\begin{equation}
\left( M^{AB}_c\left( W \right) \right)_{j, p, k_j, k_p}  =  
2 \color{black} (U - {\mathbbm 1})^B_C (W^{AC})_{j, p, k_j, k_p} \ , 
\end{equation}
\begin{eqnarray}
& & \left( M^{AB}_s\left[  \Psi^*, \Psi \right] \right)_{j, p, k_j, k_p}  =  \sum_{(n_j, n_p) \in {{\mathbb Z}^2}} \exp{\left(+ i k_j n_j + i k_p n_p \right)} 
\mu_{j, n_j, p, n_p} \ ,
\end{eqnarray}
\begin{eqnarray}
\mu_{j, n_j, p, n_p} & = & - 2
(\Psi^A)^*_{j - n_j, p - n_p} \left( 
\left.  \left( \delta^B_C D_{jj}  -  U^B_C D_{pp} \right) \right) (\Psi^C) \right)_{j + n_j, p + n_p} \nonumber \\
 & + & 
 2 \left( D_{jj}(\Psi^A)^*\right)_{j - n_j, p - n_p} \left(D_j (\Psi^B) \right)_{j + n_j, p + n_p} 
 - 2 \left( D_{pp}(\Psi^A)^*\right)_{j - n_j, p - n_p} (U \sigma_3)^B_C \left(D_j (\Psi^C) \right)_{j + n_j, p + n_p}.
\color{black}
\end{eqnarray}
\color{black}

The quantity $K^{AB}\left[\Omega\right]$ involves discrete Fourier transforms of discrete time- and space-derivatives. 
To rewrite these into a more
appealing form, we use the computation carried out in the previous section.
One finds:
\color{black}
\begin{eqnarray}
\left( K^{AB}\left[ \Omega \right] \right)_{j, p, k_j, k_p} & = & \sum_{(n_j, n_p) \in {{\mathbb Z}^2}} \exp{\left(+ i  k_j n_j + i k_p n_p\right)} 
 \times  
\left( - i (\sin k_j \color{black}) \delta^B_C + i (\sin k_p \color{black}) (U \sigma_3)^B_C  \right) \Omega^{AC}_{j, p, n_j, n_p}   \nonumber \\
& = & - i \left( (\sin k_j \color{black}) \delta^B_C - (\sin k_p \color{black}) (U \sigma_3)^B_C  \right) W^{AC}_{j, p, k_j, k_p} \nonumber \\
& = & \left( K^{AB}\left( W \right) \right)_{j, p, k_j, k_p}.
\end{eqnarray}

The final form of the discrete evolution equation obeyed by $W$ is thus:
\begin{equation}
\left(  \delta^B_C D_j   - (U \sigma_3)^B_C D_p \right) W^{AC} -  K^{AB}\left( W\right)  -  M^{AB}_c\left( W\right)  =
 M^{AB}_s \left[ \Psi^*, \Psi \right].
\end{equation}

In contrast with the transport equation obeyed by the Wigner function of the continuous Dirac field,  this equation does not involve only $W$ but also a functional of $\Psi$, $\Psi^*$. 
In the discrete case, this equation should therefore be viewed primarily as a discrete integro-differential equation obeyed by $\Psi$ and $\Psi^*$, not $W$. 
The lowest order correction to the continuous limit
is presented in the next section as an example.

\subsection{Corrections to the continuous transport equation}

Let us show how to perform an expansion around the continuous limit by computing the lowest order corrections to the transport equation induced by the discreteness of the
space-time lattice.

At second order in $\epsilon$, the operator $U(\theta)$ reads
\begin{equation}
U(\theta) \sim 
\left( \begin{array}{cc}  1 - \frac{\epsilon^2 m^2}{2} & -i \epsilon m  \\
- i \epsilon m &  1 - \frac{\epsilon^2 m^2}{2}  \end{array} \right),
\end{equation}
where $\sim$ designates equality at second order.
Since 
$D_p = \epsilon \partial_x$, 
one finds that 
\begin{equation}
U(\theta) \sigma_3 D_p \sim \epsilon
\left( \begin{array}{cc}  1 & + i \epsilon m  \\
- i \epsilon m &  - 1 \end{array} \right) \partial_x,
\end{equation}
leading to 
\begin{equation}
 \delta^A_B D_j   - (U \sigma_3)^A_B D_p  \sim \epsilon \left( 
 \delta^A_B \partial_t  - (\sigma_3)^A_B \partial_x \right) - \epsilon^2 m \, (\sigma_2)^A_B \partial_x
\end{equation}
where $\sigma_2$ is the second Pauli matrix.

Since $k_j = \epsilon k_t$ and $k_p = \epsilon k_x$,
\begin{eqnarray}
K^{AB}\left( W \right)  \sim - i \epsilon \left( k_j \delta^B_C- k_p (\sigma_3)^B_C + \epsilon m k_p (\sigma_2)^B_C 
\right) W^{AC}.
\end{eqnarray}
The first part of the mass term reads
\begin{eqnarray}
M^{AB}_c\left( W \right)  \sim - i \epsilon m (\sigma_1)^B_C W^{AC} - \frac{\epsilon^2 m^2}{2} W^{AB}.
\end{eqnarray}

The other part of the mass term depends linearly on the second derivatives $D_{jj}$ and $D_{pp}$. To compute these at second order in $\epsilon$, one can use the first order
expressions of $D_j$ and $D_p$, which are $D_j \sim \epsilon \partial_j$ and $D_p \sim \epsilon \partial_p$, combined with the zeroth order expression for $U$, which is $U \sim 
{\mathbbm 1}$. Using then the Dirac equation to eliminate the second derivatives leads to:
\begin{eqnarray}
M^{AB}_s\left[ \Psi^*, \Psi \right] 
\sim  + \color{black} 2 \epsilon^2 m^2  W^{AB}.
\end{eqnarray}
Including the lowest order corrections to the continuous case, the transport equation obeyed by $W$ thus reads:
\begin{eqnarray}
\left[(\partial_t + i k_j) \delta^B_C -(\partial_x + i k_p) (\sigma_3)^B_C + i m (\sigma_1)^B_C\right] W^{AC} = \nonumber \\
 \epsilon 
\left[
m  ( \partial_x - i k_p) (\sigma_2)^B_C + \frac{3m^2}{2} \delta^B_C
\right]W^{AC}.
\end{eqnarray}
\color{black}

\section{Conclusion}

We have defined a relativistic Wigner function for free DTQWs on the square $2D$ space-time lattice. This definition uses the concepts of discrete derivatives and distributions which 
we have also introduced. We have established the transport equation obeyed by the relativistic Wigner function and proved that this equation degenerates at the continuous limit into the transport equation obeyed by the Wigner function of Dirac fermions. We have finally computed the first corrections to this equation induced by the discreteness of the lattice.

These results can be extended in several directions. One should first address DTQWs defined, both on higher dimensional and on more general lattices, like for example planar \color{black} triangular and hexagonal \color{black} ones  \cite{ADMMP18a,JDW19a}. This could be done by using 
Fourier series defined on general spectral sets (see for example \cite{X10a} and references therein). \color{black} An extension to DTQWs defined on graphs should also be envisaged 
\cite{BdVDMPRS16a,DAEPT16a}.
\color{black}
Since several DTQWs with non constant mixing operators can be interpreted as fermions coupled to discrete gauge fields \cite{CW13a,DBD13a,DBD14a,AFF16a,Bru16a,AF16a,AP16a,AF16b,AD17a,AMBD16a,CGW19a}\color{black}, one should define a Wigner function which incorporates these gauge fields and, in particular the electromagnetic field and the gravitational field. For example, defining a gauge-invariant Wigner function for Dirac particles couples to electromagnetic fields is highly non trivial \cite{Vasak87a} and one wonders how the problem translates to DTQWs. Finally, DTQWs defined through unitaries which present a time randomness decohere and behave asymptotically like non quantum diffusions (see for example \cite{MD16a} and references therein). One then expects the transport equation for the relativistic Wigner function to approach asymptotically relativistic transport equations similar to those obtained for relativistic stochastic processes \cite{DMR97a,CD08a}. This should be confirmed and the asymptotic fully analyzed. 

%
%
%

%
%

\vspace{6pt}

\end{document}